\documentclass[12pt]{article}
\def\pd{\partial}
\def\be{\begin{equation}}
\def\ee{\end{equation}}
\def\bea{\begin{eqnarray}}
\def\eea{\end{eqnarray}}
\begin{document}
05.09.04
\begin{center}
\begin{Large}
{\bf Implicit solutions to some Lorentz invariant non-linear equations revisited.}
\end{Large}
\\
\vspace{0.5cm}
{\bf David B. Fairlie
$\footnote{e-mail: David.Fairlie@durham.ac.uk}$ 
}\\
\vspace{0.3cm}
Department of Mathematical Sciences,\\
         Science Laboratories,\\
         University of Durham,\\
         Durham, DH1 3LE, England \\
\end{center}

\vspace{0.2cm}
\begin{abstract} An implicit solution to the vanishing of the so-called Universal Field Equation, or Bordered Hessian, which dates at least as far back as 1935 \cite{chaundy} is revived, and 
derived from  a much later form of the solution.  A linear ansatz for an implicit solution of second order partial differential equations, previously shown to have wide applicability \cite{fai} is at the heart of the Chaundy solution, and is shown to yield solutions even to the linear wave equation.

\end{abstract}
\section{Introduction}
Some years ago, the notion of  a Universal Field Equation, so called because it admits an infinite number of 
inequivalent Lagrangians \cite{gov} was introduced. Such equations arise in several ways; in 3 dimensions as the condition for developable surfaces \cite{eis}, or more generally as a measure of the total curvature of a surface $\phi(x,y,z)\,=\,0$ \cite{chaundy}, or as a consequence of the the hydrodynamic  equations for vanishing total rate of change of velocity \cite{dbfint}, i.e. 
\[\frac{D}{dt}\vec v \,=\,\frac{\pd}{\pd t}\vec v + \vec v\cdot\nabla \vec v \,=\,0.\] This last approach is sketched in the first section where it is illustrated for the case of a two component vector field. If $u(t,x,y)$ and $v(t,x,y)$ are expressible  in terms of a single function $\phi(t,x,y)$, then it is shown that the consequence is the Universal Field Equation, which Chaundy more prosaically calls the bordered Hessian;
\begin{equation}
\det\left|\begin{array}{cccc}
          0&\phi_t&\phi_x&\phi_y\\
       \phi_t&\phi_{tt}&\phi_{tx}&\phi_{ty}\\                    
        \phi_x&\phi_{tx}&\phi_{xx}&\phi_{xy}\\
        \phi_y&\phi_{ty}&\phi_{xy}&\phi_{yy}
\end{array}      \right|\,=\,0.\label{bat3}
\end{equation}
This equation is a natural generalisation of the Bateman equation;
\[\phi_{tt}\phi_x^2 -2\phi_{tx}\phi_t\phi_x+ \phi_{xx}\phi_t^2\,=\,0\]
about which much is known \cite{gov}. In particular, the general solution to this Bateman equation is given by constraining two arbitrary functions of $\phi$; say $f_1(\phi)$ and $f_2(\phi)$  by the relation
\[tf_1(\phi)+xf_2(\phi)\,=\,1.\]
The implicit solution of this equation gives the general solution (dependent upon two arbitrary functions, as expected for a second order partial differential equation) to the Bateman equation. The ubiquity of this linear construction of an implicit solution to a second order partial differential equation, which has been explored in \cite{fai}, is further investigated here for not only equations of the form (\ref{bat3}), but even the linear wave equation in arbitrary dimension.
It turns out that a solution to (\ref{bat3}) has been known for some time,\cite{chaundy} and maybe earlier; in ignorance of this solution Govaerts and I found a solution by means of Legendre transforms much later \cite{fab}. There is an unpublished solution on similar lines to the solution of the Monge-Amp\`ere equation in \cite{leznov}. The first two methods of solution are explained and related in the third section, which gives the most direct proof of the solution. The main purpose of this paper is to publicise Chaundy's forgotten solution and
on the way collect some results on the two dimensional nonlinear wave equation from which the 
Universal Field Equation arises as a second order equation. 

\section{Two fields-first order}
This section reviews the simplest case beyond that of Bateman which occurs for a two component velocity field $(u,\ v)$ in one time, two space dimensions.
 The first order
equations  (sometimes called after Monge) are 
\begin{eqnarray}
\frac{\partial u}{\partial t}&=& u\frac{\partial u}{\partial x}+
v\frac{\partial u}{\partial y},\label{one}\\
 \frac{\partial v}{\partial t}&=& u\frac{\partial v}{\partial x}+
v\frac{\partial v}{\partial y}.\label{two}
\end{eqnarray}
These equations admit general solutions of the form;
\[
 u\,=\, F(x-ut,y-vt);\ \ \ v\,=\, G(x-ut,y-vt)
\]
in terms of two arbitrary functions, $F,\ G$. \cite{leznov}\cite{fai1}
This is easy to verify; the easiest way to find a constructive derivation is to
change the nature of the independent variables; instead of thinking of $(u,\ v)$ as given in terms of the variables $(t,\ x,\ y)$, think instead of regarding  
$(x,\ y)$ as functions of $(t,u,v)$. Then the equations of transformation are;
\begin{eqnarray}
\frac{\partial u}{\partial t} &=&\frac{\frac{\partial x}{\partial t}\frac{\partial y}{\partial v}-\frac{\partial x}{\partial v}\frac{\partial y}{\partial t}}{\frac{\partial x}{\partial u}\frac{\partial y}{\partial v}-\frac{\partial x}{\partial v}\frac{\partial y}{\partial u}},\nonumber\\
\frac{\partial u}{\partial x}&=&\frac{\frac{\partial y}{\partial v}}{\frac{\partial x}{\partial u}\frac{\partial y}{\partial v}-\frac{\partial x}{\partial v}\frac{\partial y}{\partial u}},\nonumber\\
\frac{\partial u}{\partial y} &=&-\frac{\frac{\partial x}{\partial v}}{\frac{\partial x}{\partial u}\frac{\partial y}{\partial v}-\frac{\partial x}{\partial v}\frac{\partial y}{\partial u}},\nonumber
\end{eqnarray}
with similar equations for derivatives of $v$.
In terms of the interchanged derivatives, the equations (\ref{one}, \ref{two})
become
\begin{eqnarray}
\left(\frac{\partial x}{\partial t}-u\right)\frac{\partial y}{\partial v}-\left(\frac{\partial y}{\partial t}-v\right)\frac{\partial x}{\partial v}&=&0,\nonumber\\
\left(\frac{\partial x}{\partial t}-u\right)\frac{\partial y}{\partial u}-\left(\frac{\partial y}{\partial t}-v\right)\frac{\partial x}{\partial u}&=&0.\nonumber
\end{eqnarray}
These equations, for a non trivial solution yield
\begin{equation}
x-ut \,=\,f(u,v),\\\ y-vt \,=\,g(u,v).
\end{equation}
These solutions are equivalent to the general solution given above.
The generalisation to $n$ equations is immediate;
We have equations of the type
\begin{equation}
\det\left|\begin{array}{cccc}
\frac{\pd x_1}{\pd t}-u_1&\frac{\pd x_2}{\pd t}-u_2&
\cdots&\frac{\pd x_n}{\pd t}-u_n\\
\frac{\pd x_1}{\pd u_2}&\frac{\pd x_2}{\pd u_2}
&\cdots&\frac{\pd x_n}{\pd u_2}\\
\vdots&  \vdots&\ddots&\vdots\\
\frac{\pd x_1}{\pd u_n}&\frac{\pd x_2}{\pd u_n}&\cdots&\frac{\pd x_n}{\pd u_n}
\end{array}\right|\,=\,0,\label{det} 
\end{equation}
together with similar equations where each subsequent row of 
$\displaystyle{\det\left|  \frac{\pd x_j}{\pd u_k}\right|}$ is replaced in turn by the first row of the above determinant (\ref{det}). 
There exists an infinite number of Lagrangians for (\ref{one},\ \ref{two}). 
\begin{equation}
{\cal L}\,=\, \frac{\partial A(u,v)}{\partial v}\frac{\partial u}{\partial t}+
u\frac{\partial A(u,v)}{\partial v}\frac{\partial u}{\partial x}+
 \left(v\frac{\partial A(u,v))}{\partial v}-A(u,v)\right)\frac{\partial u}{\partial y}.
\end{equation}
Here $A(u,v)$ is an arbitrary function of $(u,\ v)$. 
If $u(\phi)$ and $v(\phi)$ are functions of a single argument $\phi(x,y,t)$
then you can permute derivatives
\[\frac{\partial u}{\partial x}\frac{\partial v}{\partial y}
\,=\,\frac{\partial u}{\partial y}\frac{\partial v}{\partial x}.\]
This leads  to the Universal Field Equation for $\phi$ (\ref{bat3}), since the two Monge equations become the single equation
\be
\frac{\partial \phi}{\partial t}+\frac{\partial u}{\partial \phi}\frac{\partial \phi}{\partial x}+\frac{\partial v}{\partial \phi}\frac{\partial \phi}{\partial y}\,=0.
\ee
Forming the eliminant of this equation, together with its derivatives with respect to $(t,\ x,\ y)$ gives the determinantal equation (\ref{bat3}).

\section{Chaundy's solution}
Consider four arbitrary functions of two variables, $(\phi,u)$ denoted by $F_i(\phi,u),\ i= 1\dots 4.$
Actually the fourth function is redundant and may be taken as 1. 
These functions are subject to the following constraints;
\bea
tF_1(\phi,u)+xF_2(\phi,u)+yF_3(\phi,u)&=&F_4(\phi,u),\label{cho1}\\
t\frac{\pd}{\pd u}F_1(\phi,u)+x\frac{\pd}{\pd u}F_2(\phi,u)+y\frac{\pd}{\pd u}F_3(\phi,u)&=&\frac{\pd}{\pd u}F_4(\phi,u).\label{cho2}
\eea
The first equation represents a two parameter family of planes; this together with the second ,
upon elimination of the parameter $u$ gives the envelope of these planes; hence the connection with the theory of developable surfaces as presented in \cite{eis}. Notice that in fact {\ref{cho1}} and indeed (\ref{cho2} are of the form of the universal solution, (\cite{fai}), the linear ansatz which seems applicable to many second order partial differential equations.
This solution may be verified in the following manner; calculate the  first and second derivatives ;  setting
\bea\mu &=&t\frac{\pd F_1}{\pd \phi}+x\frac{\pd F_2}{\pd \phi}+y\frac{\pd F_3}{\pd \phi}-\frac{\pd F_4}{\pd\phi}.
\nonumber\\
\lambda &=&t\frac{\pd^2 F_1}{\pd \phi^2}+x\frac{\pd^2 F_2}{\pd \phi^2}+y\frac{\pd^2 F_3}{\pd \phi^2}-\frac{\pd^2 F_4}{\pd \phi^2}.\nonumber
\eea
typical derivatives are given by
\bea
\phi_t &=&\frac{-F_1}{\mu}\nonumber\\
\phi_{tt}&=&\frac{1}{\mu}\left(-2\frac{\pd F_1}{\pd \phi}\phi_t-(\phi_t)^2\lambda -\frac{\pd F_1}{\pd u}u_t\right),\nonumber\\
\phi_{tx}&=&\frac{1}{\mu}\left(-\frac{\pd F_1}{\pd \phi}\phi_x-\frac{\pd F_2}{\pd \phi}\phi_t-\phi_t\phi_x\lambda -\frac{\pd F_1}{\pd u}u_x\right).\nonumber
\eea

Consistency with the alternative construction of $\phi_{tx}$ by first differentiating with respect to $x$ requires
\[\frac{\pd F_1}{\pd u}\frac{\pd u}{\pd x}\equiv\frac{\pd F_2}{\pd u}\frac{\pd u}{\pd t}\]
etc. Substitution into (\ref{bat3}) together with some elementary row and column operations shows that the determinant indeed vanishes. Since there are initially four arbitrary functions, one of which is redundant, and
another is involved in the second constraint, two remain arbitrary, as is required by the general solution of 
a second order partial differential equation.
 
\section{Linearision by Legendre Transforms}
   The Legendre Transform  was employed in \cite{fab} to linearize the Universal Field Equation  This transform, which is clearly involutive, has the flavour of a twistor transform. In a multivariable generalisation  the Universal Field Equation runs as  follows:
\be
\frac {\partial \xi_{\alpha}}{\partial x_n}+\sum_{\beta=1}^{n-1}
\xi_{\beta} \frac {\partial \xi_{\alpha}}{\partial x_{\beta}}=0 \label{univ}
\end{equation}

Introduce a dual space with co-ordinates $\xi_i,\ i=1,\dots,d$ and a function
$w(\xi_i)$ defined by
\bea
\phi(x_1,x_2,\dots,x_d)+w(\xi_1,\xi_2,\dots,\xi_d)=
&x_1\xi_1+x_2\xi_2+\dots ,x_d\xi_d.\\
\xi_i={\pd\phi\over\pd x_i},\quad x_i={\pd w\over\pd \xi_i},\quad& \forall{i}.
\eea
To evaluate the second derivatives $\phi_{ij}$ in terms of derivatives of $w$
it is convenient to introduce two Hessian matrices;
$\Phi,\  W$ with matrix elements  $\phi_{ij}$ and $w_{\xi_i\xi_j}=w_{ij} $
respectively. Then assuming that $\Phi$ is invertible, $\Phi W=1\!\!1$  
and
\be
 {\pd\sp2\phi\over\pd x_i\pd x_j}= ( W\sp{-1})_{ij},\quad
 {\pd\sp2w\over\pd \xi_i\pd \xi_j}= (\Phi\sp{-1})_{ij}.
\ee

 The effect of the Legendre transformation upon the equation (\ref{univ}) is immediate; in the new variables the equation becomes simply
\be
\sum_{i,j}\xi_i\xi_j \frac{\pd^2 w}{\pd \xi_i\pd \xi_j}=0.
\label{univ3}
\ee
a linear second order equation for $w$. The general solution of this equation is immediate; $w=f_1(\xi_1\dots\xi_d)+f_2(\xi_1\dots\xi_d)$ where $f_0$ is an arbitrary homogeneous function of weight zero and $f_1$ is an arbitrary homogeneous function of weight one. Then the general solution of the Universal Field Equation is given implicitly as the eliminant of the variables $\xi_i$ from the equations
\be
\phi(x_1,x_2,\dots,x_d)=\sum x_j\xi_j-w(\xi_1,\xi_2,\dots,\xi_d)
,\quad x_i={\pd w\over\pd \xi_i}.
\label {useful}
\ee
This procedure may be used to verify the general solution of the Bateman equation.\cite{fab}
Note that (\ref{useful}) implies that 
\be
\sum x_i\frac{\pd\phi}{\pd x_i}=f_1(\xi_i).
\ee
Hence if $f_1=0$ then $\phi$ is an arbitrary homogeneous function of degree zero in its arguments.
How does does this solution equate to Chaundy's?
By differentiating $w$ with respect to $\xi_i$
\be
 x_i\,=\, \frac{\pd f_0}{\pd\xi_i}+\frac{\pd f_1}{\pd\xi_i}.
\ee
Multiplying by $\xi_i$ and summing, having regard to the homogeneities;
\be
\sum x_i\xi_i\,=\,f_1;\ \ \ \phi(x_1,x_2,\dots x_d).\,=\,f_0.\label{hom}
\ee
The crucial observation is that each $\xi$ may be regarded as a function of $d-1$ arguments as a consequence of the homogeneities, i.e.
\[ \xi_j = \xi_j(\phi,u_k),\ \ \ k=1\dots d-2.\]
Additionally, from the second member of (\ref{hom}), 
\[\frac{\pd f_0}{\pd u_k}\,=\,0.\]
Thus from (\ref{useful}),
\bea
\sum_j x_j\xi_j(\phi,\xi_i) &=&f_1(\phi,u_i),\nonumber\\
\sum_jx_j\frac{\pd\xi_j}{\pd u_k}& =&\frac{\pd f_1}{\pd u_k}.\nonumber
\eea
These are precisely the same as Chaundy's constraints. The elimination of the variables $u_k$ from this set, 
provides an implicit solution of the universal field equation, or bordered Hessian.
\section{Monge Amp\`ere}
A similar result can be proved for the homogeneous Monge-Amp\`ere , or vanishing Hessian equation;
\be
\det\left|\frac{\pd\phi}{\pd x_i\pd x_j}\right| \,=\,0.\label{ma}
\ee
In fact, as noted in \cite{chaundy}, there is a close connection between this problem and the universal field equation;
in the case of n variables $x_i,\ i\,=\,1\dots n$,
if $\phi(x_i)\,=\,0$ is solved implicitly for $x_n\equiv z = z(x_j),\ j\,=\,1\dots n-1$,
then
\be
\det\left|\frac{\pd z}{\pd x_i\pd x_j}\right| \,=\,\frac{1}{\phi^{n+1}}
\det\left|\begin{array}{cc}
0&\frac{\pd \phi}{\pd x_i}\\
\frac{\pd \phi}{\pd x_j}&\frac{\pd^2 \phi}{\pd x_ix_j}
\end{array}\right|\label{eqiv} 
\end{equation}

In the case of three independent variables, $(x_1,\ x_2,\ x_3)$ the equation is given,\cite{chaundy} by
\bea
\phi(x_1,x_2,x_3)&=& x_1G_1(u,v)+x_2G_2(u,v)+x_3G_3(u,v)-G_4(u,v)\nonumber\\
\frac{\pd}{\pd u}G_4(u,v)&=&x_1\frac{\pd}{\pd u}G_1(u,v)+x_2\frac{\pd}{\pd u}G_2(u,v)+x_3\frac{\pd}{\pd u}G_3(u,v)\label{one1}\\
\frac{\pd}{\pd v}G_4(u,v)&=&x_1\frac{\pd}{\pd v}G_1(u,v)+x_2\frac{\pd}{\pd v}G_2(u,v)+x_3\frac{\pd}{\pd v}G_3(u,v)\label{two1}
\eea
From this it is easy to calculate second derivatives of $\phi$,
\be
\phi_{x_ix_j}\,=\, \frac{\pd G_i}{\pd u}u_{x_j}+\frac{\pd G_i}{\pd v}v_{x_j}+\frac{\pd G_j}{\pd u}u_{x_i}+\frac{\pd G_j}{\pd v}v_{x_i}\label{three}
\ee
as a consequence of (\ref{one1}) and (\ref{two1}).
From this representation of the second derivative it is easy to see that the Hessian (\ref{ma}) vanishes, and the principle of construction can be easily inferred for the $n\times n$ case. A general solution to this equation has also been given by Leznov and myself \cite{leznov} 
Note that
\be
\sum_{i=1}^{i=3} \phi_{x_i}x_i = \sum_{i=1}^{i=3}G_ix_i\,=\,\phi(x_1,x_2,x_3)+G_4(u,v)
\ee
so if $G_4(u,v)=0$, then $\phi$ is homogeneous of weight one, a well known solution which makes the Hessian vanish. 
\section{The linear wave equation}
Surprisingly, the wave equation in $n$ space dimensions,
\be
\frac{\pd^2 u}{\pd t^2}- \sum_{i=1}^{i=n}\frac{\pd^2 u}{\pd x_i^2} \,=\,0\label{linear}
\ee
also yields to a solution in terms of the linear ansatz
\be
 tF_0(u) + \sum_{i=1}^{i=n}x_iF_i(u)\,=\,1\label{ans}
\ee
where the coefficients are arbitrary functions of $\phi$. Denoting $t$ by $x_0$, the wave equation becomes
\be
-\frac{ \left(\sum_{j=1}^{j=n} x_jF''_j\right)\left(F_0^2-\sum_{k=1}^{k=n}F_k^2\right)}
{\left(\sum_{j=0}^{j=n}x_jF'_j\right)^3} +\frac{2\left(F'_0F_0-\sum_{j=1}^{j=n}F'_jF_j\right)}{\left(\sum_{j=0}^{j=n}x_jF'_j\right)^2}\,=\,0,
\ee
and primes denote, as usual, differentiation with respect to the argument $u$. This equation will be satisfied provided the functions $F_j$ are subject to the single constraint
\be
F(u)_0^2-\sum_{k=1}^{k=n}F(u)_k^2\,=\,0. \label{ans2}
\ee
That such a constraint is necessary may be seen from the observation that  while the ansatz and the universal equations are form invariant under a functional change in $\phi$ or $u$, the wave equation does not possess this property. In fact, the solutions obtained by this method also satisfy the null wave vector identity
\be
\left(\frac{\pd u}{\pd x_0}\right)^2-\sum_{k=1}^{k=n}\left(\frac{\pd u}{\pd x_i}\right)^2\,=\,0. \label{ans3}
\ee
and are thus not the most general solution. However, the general solution may be constructed from a superposition of such solutions by invoking the additive principle of solutions to linear equations; eg in 2+1 dimensions we may take $F(u)_0 =\frac{1}{f^{-1}(u)},\ F(u)_1 =\frac{1}{f^{-1}(u)}\cos(\theta),\ F(u)_2 =\frac{1}{f^{-1}(u)}\sin(\theta)$.
Then 
\be
u\,= f(x_0+x_1\cos(\theta)+x_2\sin(\theta),\theta)\label{sol1}
\ee
is the standard solution to the wave equation, for an arbitrary function $f$, which may also depend upon $\theta$ and the general solution is obtained from this by integration over $\theta$ \cite{WW}

\section{Conclusion}
The linear ansatz, (\ref{ans}),  has been shown to provide a universal framework in seeking solutions to 
partial differential equations of second order, even including the linear case. A feature of the equations studied here is that they are all invariant under linear (Lorentz) transformations of the independent variables, but this is not essential \cite{fai}


\begin{thebibliography}{}
\bibitem{chaundy}T. Chaundy {\it The Differential Calculus} The Clarendon Press, Oxford  (1935)
\bibitem{fai} D.B. Fairlie, A Universal Solution {\it J. Nonlin. Math. Phys}{\bf 9} (2002) 256-261.
\bibitem{gov} D. B. Fairlie, J. Govaerts and A. Morozov, Universal Field Equations with Covariant Solutions,
{\it Nucl. Phys.} {\bf B373} (1992) 214-232.

\bibitem{eis}L.P. Eisenhart, {\it An Introduction to Differential Geometry} Princeton Umiversity Press,  (1940) 54-59.
\bibitem{leznov}D.B. Fairlie and A.N. Leznov, General solutions of the  Monge-
Amp\`{e}re equation in $n$-dimensional space,{\it Journal of Geometry and Physics}.
{\bf 16} (1995) 385-390.{\bf hep-th/9403134}
\bibitem{fai1}
D.B. Fairlie,Formal Solutions of an Evolution Equation of Riemann type,
{\it Studies in Applied Math} {\bf 98} (1997) 203-205.

\bibitem{dbfint} D.B. Fairlie, Integrable Systems in Higher Dimensions 
{\it Quantum Field Theory, Integrable Models and Beyond} 
Editors. T. Inami  and R. Sasaki  {\it Progress of Theoretical Physics Supplement}
 {\bf 118} (1995) 309-327.

\bibitem{fab} D. B. Fairlie and J. Govaerts, Linearization of  Universal Field Equations, {\it J. Phys.}
{\bf A 26} (1993) 3339-3347.
\bibitem{WW}  E.T.Whittaker and G.N. Watson {\it Modern Analysis} Cambridge University Press (1963) p390.
 
\end{thebibliography}
\end{document}